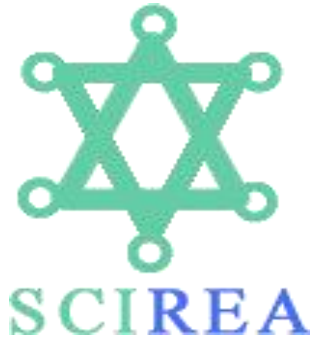

# Implications of detecting the Axion-Like Particles in Astro-Particle Physics

K K Singh[*]

Astrophysical Sciences Division, Bhabha Atomic Research Centre, Mumbai- 400085 India

([*]E-mail: kksastro@barc.gov.in)

## Abstract

Axions and axion like particles in general are consequences of the extensions of the standard model in particle physics. Axions have been proposed as hypothetical pseudo-scalar particles in particle physics to solve the strong Charge-Parity problem using quantum chromodynamics. These particles are also proposed to be an attractive candidate for cold dark matter in the Universe. The current understanding of the existence and properties of the axions is not very clear and these hypothetical particles remain elusive. However many theoretical and experimental searches are being made for understanding and constraining the properties of axions via detection of axion like particles. The term axion like particles is used to include different types of axions going beyond the standard model. In this paper, we review the different physical properties of axions and axion like particles and their possible sources in the laboratory and Universe. We discuss different experimental set ups for detection of axion like particles proposed in the literature. Implications of detecting the axion like particles in the areas of particle physics, astrophysics and cosmology are also discussed in the framework of present and future motivations.

**Keywords :** Axions- elementary particles: ALPs- astroparticle physics: detections

# 1 Introduction

Axion represents a special class of hypothetical particles beyond the standard model (SM) in particle physics. It was proposed in 1977 by Peccei and Quinn (PQ) to solve the charge-parity (CP) problem in SM using quantum-chromodynamics (QCD) and for associating the particles with axial anomaly in strong interactions [1, 2]. The QCD Lagrangian has an interaction term which violates CP symmetry and its effect is characterized by a dimensionless parameter with very small numerical value. The CP violation is not observed in strong interactions due to spontaneous breaking of PQ symmetry which leads to the creation of axions in QCD. The axions in QCD are pseudo-Goldstone bosons and are assumed to acquire mass from the breaking of PQ chiral symmetry like the Higgs bosons which gain mass from the spontaneous symmetry breaking of scalar field in quantum field theory. Apart from solving the strong CP problem in SM of particle physics, axions are thought to play very important role in astrophysics and cosmology. The QCD axions are hypothesised to be strong candidate for dark matter at the PQ symmetry breaking energy scale [3]. The dark matter is an outstanding mystery toady in particle physics, astrophysics and cosmology. The interface between these three branches of science is generally referred to as astro-particle physics. The exact nature of dark matter cannot be fully explained by the SM of particle physics and therefore new physics beyond the SM needs to be explored. In string theory, axions represent the zero modes of the pseudo-scalar fields associated with the compact spatial dimensions. Many new ideas have been proposed for the direct detection of axions in laboratory as well as astrophysical conditions in the Universe. In the following Sections, we qualitatively discuss different aspects related to the observation of special class of hypothetical axions known as Axion Like Particles (ALPs) in astro-particle physics. In Section 2, we have briefly described the concept of ALPs and their properties which are known so far. In Section 3, different types of experiments proposed for detecting axions and ALPs are discussed. In Section 4, implications of axions and ALPs in particle physics, astrophysics and cosmology are outlined. We have presented the summary in Section 5.

# 2 Axion Like Particles

Axion-Like Particles (ALPs) are a generalization of QCD axion from the extensions of SM in particle physics. These are predicted to arise in string theory compactification. The scalar ALPs are produced by the photons with polarization vector. Therefore, ALPs strongly couple to photons in the presence of external magnetic field. The QCD axions are mainly characterized by low mass and weak coupling coming from the spontaneous PQ symmetry breaking at ultra-high energies. Whereas ALPs are independent of PQ mechanism and acquire mass from the dynamics of the global symmetry breaking. ALPs can also be produced thermally in the hot and dense astrophysical environments due to their coupling to the particles in the SM like electrons, nucleons and photons.

## 2.1 Axions and ALPs

The conversion between QCD axions and ALPs can take place adiabatically in the physical conditions equivalent to the early Universe due to the existence of hypothetical mass mixing similar to the neutrino oscillations. This adiabatic conversion of QCD axions into ALPs is found to be a possible explanation for dark matter abundance in the present day Universe. The ALPs represent generalized QCD axions and are proposed to be low energy consequences of fundamental physics. The cosmological dynamics for axions and ALPs are similar. If the mass and decay constant of ALPs are smaller than that for QCD axions at zero temperature, level crossing takes place between the two. The level crossing is a phenomena in which adiabatic conversion of QCD axions into ALPs and vice versa takes place due to mass eigenstates similar to the neutrino oscillations into different flavours [4,5]. The axion abundance is reduced by the adiabatic conversion due to existence of mass mixing eigenstates. ALPs are also strongly motivated candidates for cold dark matter just like axions in the Universe. The photophobia is assumed to be a natural property of axions due to which the coupling to photons is suppressed if the underlying parameters are not properly defined. It means the coupling of photophobic axions to SM boson do not involve photons. The signature of axions with photophobia is different from the typical ALPs and they are important is collider-based experiments for probing higher mass axions.

## 2.2 Physical Properties

The axions are Nambu-Goldstone bosons of spin zero associated with PQ symmetry in gauge theory. These particles are associated with a real scalar field. The axions are denoted by the symbol **a** and their properties are characterized by the single parameter called decay constant ($f_a$) which defines the energy scale of axion interaction or spontaneous breaking of PQ symmetry. The mass of axion ($m_a$) and coupling strength ($g_a$) both are inversely proportional to the decay constant. This means for lighter mass axions couplings are weaker. The value of axion decay constant has been constrained to be $10^9$ GeV $\leq f_a \leq 10^{12}$ GeV from the studies based on the astrophysics of axions involving cosmology and stellar evolutions [6]. At this energy scale, the PQ symmetry is expected to be spontaneously broken in the strong interactions. This higher energy scale implies that axions are low mass weakly interacting particles. The upper and lower bounds on the decay constant give mass range of $10^{-6}$ eV $\leq m_a \leq 10^{-3}$ eV and coupling constant limits of $10^{-15}$ GeV$^{-1}$ $\leq g_a \leq 10^{-11}$ GeV$^{-1}$ for QCD axions. With this parameter space, axions are assumed to be possible candidates for cold dark matter (CDM) in cosmology and weakly interacting with ordinary matter particles. In general, ALPs only interact with the photons via electromagnetic coupling whereas QCD axions couple to the strong forces as well as photons. For ALPs, the coupling constant and mass are independent of each other and therefore their parameter space is not well constrained as compared to QCD axions. The allowed mass range for ALPs is $10^{-33}$ eV $\leq m_{ALP} \leq 10^{-18}$ eV [7]. The QCD axions are coupled to the standard model whereas ALPs are weakly coupled to the standard model. Axions should also interact gravitationally since they acquire mass but their coupling to gravity is minimal.

**2.3 Photon-ALP Conversion**

Photons are converted to ALPs in the coherent magnetic field transverse to their direction of propagation [8, 9]. Similarly, axions or ALPs couple to photons via two photon coupling in the presence of background magnetic field. Solution of equation of motion for photon-ALP conversion (or vice versa) suggests that the conversion probability is maximum and independent of the energy of particles above a critical energy. This critical energy is inversely proportional to the strength of external magnetic field. If the polarization vector of the photons is parallel to the external transverse magnetic field, pseudo-scalar ALPs are induced whereas scalar ALPs are produced if the two quantities are perpendicular. This means photon-ALP conversion behaves like a polarimeter because it changes the polarization states of photons. The coherent photon-ALP conversion is an oscillation process similar to the neutrino oscillation. ALPs are spin zero pseudo-scalar bosons whereas photons are particles with spin one. The spin mismatch in the photon-ALP process is compensated by the external magnetic field [9]. The most reliable theoretical explanation for photon-ALP conversion is given by the so called Primakoff effect [10]. In Primakoff effect, a photon scatters on a charged particle and converts into ALP upon exchanging a virtual photon.

**2.4 Astrophysical or Natural Sources**

Astrophysical observations have provided stringent constraints on the parameters of axions but they still lack detection of axions. ALPs can be thermally produced in hot and dense in astrophysical environments like Sun, supernova, stars and globular clusters [7]. The observed properties of these astrophysical objects can be used to search for the existence of stellar axions or ALPs. The emission rate of stellar axions can be low due to their weak coupling but emission takes place from the whole volume of the stellar source. The production of solar axions occurs till the rest mass energy ($m_a c^2$) of axions is less than core temperature (1 keV). The maximum luminosity of solar axions is observed at 3 keV. The Sun consumes more nuclear fuel due to the emission of axions. The luminosity of axions emitted from the Sun is about 4% of the solar luminosity. The flux of solar axions is described by thermal distribution with mean energy of 4.2 keV [11]. Axions can also be emitted from the stars due to their low masses. The stellar cores with hot and dense plasma are ideal sites for production of ALPs through Primakoff process [10] involving thermal photons. The ALPs produced in stellar cores escape with high energy due to their large mean free path as compared to the stellar radius. Axion emission from stars helps in understanding their evolutionary history is astrophysics. Axion stars are also proposed from the solution of equations of motion of collisionless and coherently oscillating axions [12]. The formation of axion stars and wave like nature at de Broglie wavelength scale are possible for light axions with mass **$m_a$** ~ $10^{-22}$ eV [13].

# 3 Experiments for detection of ALPs

The detection of axions or ALPs in the laboratory experiment will be able to solve the questions related to their real existence in the nature. ALPs can be detected through their conversion into photons in the presence of external magnetic field. The experimental set ups or techniques proposed for the detection of axions or ALPs are very different from the particle physics which require accelerator based high energy interactions. The lack of particular relationship between mass and decay constants of ALPs provides unique opportunity to search them in natural sources or in laboratory. Several hypothetical experiments have been proposed for the direct detection of the particles belonging to the axion class from laboratory to the astrophysical environments/natural sources. These experiments search for axions over a range of mass and coupling scales and which have been produced from the big bang during the early Universe or stars or even in the laboratory. The laboratory class of axions are detected using laser photons travelling in a constant and homogeneous magnetic field. The detection principles for axion or ALPs in different experiments are briefly described below.

## 3.1 Haloscopes

Haloscopes have been proposed to detect the axions from the dark matter galactic halos [14] and hence the name *haloscopes*. It is assumed that there are galaxies inside halo like structures formed by the dark matter particles. The density of dark matter particles at the centre of these halos is very high and therefore they decay into SM particles by scattering each other. The SM particles further decay and emit gamma-ray signal from different astrophysical objects like dwarf galaxies [15]. Haloscopes are electromagnetic cavity with arbitrary cross-sections and filled with homogeneous static magnetic field. The detection principle of conventional haloscopes is based on the fact that axions are converted into photons in the presence of external magnetic field [16]. Therefore, haloscopes are also known as cavity experiments. These experiments exploit the fact that the axions from dark matter halos are non-relativistic and monochromatic. Therefore, a resonant conversion can be made using microwave cavity with high quality factor and resonant frequency matched with the mass of axion in the magnetic field. The haloscopes are not suitable for detection of heavy axions with large masses over wide range and therefore they are useful in detecting the galactic-halo axions or cosmological axions.

## 3.2 Helioscopes

Helioscopes are telescopes with long magnet in an opaque casing to detect axions produced from the Sun [14, 16]. The solar axions can be converted to photons using inhomogeneous electric or magnetic fields and the energy photon will be equal to the energy of axions (keV) coming from the Sun. Therefore, the solar axions convert into X-ray photons after passing through the magnetic field in the helioscope at the Earth and are detected as X-ray photons by the detector placed behind the magnet. For low mass solar axions, the conversion probability is maximum in the telescope. The mass of solar axions has been constrained in the range 0.84 eV $\leq$ **$m_a$** $\leq$ 1 eV with coupling constant in the limit 5.6 ×$10^{-10}$ GeV$^{-1}$ $\leq$ **$g_a$** $\leq$ 1.4× $10^{-9}$ GeV$^{-1}$ using the helioscope detections [17]. Helioscopes using crystal detectors have also been proposed to detect the X-ray photons produced after conversion from axions [18]. The sensitivity of helioscopes with crystal detectors is independent of the axion mass. Therefore, these helioscopes can be used for detetction the axions with mass over a wide range.

### 3.3 Laser shining through a wall

This is also known as photon-regeneration experiment. This experiment [19] is based on the conversion of photons into ALPs in the presence of constant magnetic field. In this experiment, a laser beam is shined on a wall and an identical arrangement of magnets is made on both sides of the wall for constant and similar magnetic fields. A small fraction of photons in the laser beam travelling through the magnetic field is converted into axions. These axions propagate unimpeded through the wall due to their weak interaction with the ordinary matter of the wall. Some of the axions passed through the wall are reconverted to photons in the presence of similar magnetic field and can be detected in the form of dim laser light on the other side. Detection of photons behind the wall would indicate the presence of ALPs. This is also known as photon-regeneration experiment. This experiment can be performed in laboratory with good control over experimental conditions and does not depend on the sources of axions in the Universe. The only drawback with this proposed experiment is that the signal is very weak due to two stage conversion between photons and axions. These experiments are supposed to be capable of detecting ALPs with improved sensitivity in future.

### 3.4 Interferometry Experiment

Interferometry experiment has been proposed to detect ALPs in the similar way as laser shining through a wall experiment [20]. In this experimental setup, the primary laser beam is divided into two beams of similar intensity by a beam-splitter. One beam is made to travel through the region of constant magnetic field to induce photon-ALP conversion while second is used as a reference beam. After passing through the magnetic field region, the first beam is combined and made to interfere with the reference beam. The first beam is expected to have less intensity and a phase shift relative to the reference beam due to photon-ALP conversion. The measurement of change in the combined intensity will indicate the existence of ALPs in this experiment. The signal intensity at the detector is expected to be stronger due to one stage photon-ALP conversion. However, this experiment has shortcomings due to presence of shot noise occurring from the use interferometers.

### 3.5 Magneto-optical Vacuum Experiments

These experiments are based on the fact that photons disappear from the polarized laser beam passing after conversion into axions due to magneto-optical effects of vacuum [21]. When a polarized laser beam traverses through a diploe magnetic field, the photons with electric field component parallel to the external magnetic field are converted into virtual axions with small probability resulting in the rotation of polarization vector. As a result, the polarized laser beam is modified due to the vacuum birefringence. These are indirect effects to observe the existence of axions in laboratory due to magneto-optical vacuum effects. Experiments based on this effect have been able to constrain the axion mass at meV energy scale and coupling strength in the range $1.5 \times 10^{-3}$ GeV$^{-1}$ ≤ $g_a$ ≤ $8.6 \times 10^{-3}$ GeV$^{-1}$ [22].

**3.6 Astrophysical Searches**

Astrophysical environments provide coherent magnetic fields over a large distance in the Universe. Therefore, they are natural sites for observing photon-ALP conversion proceeses. Astrophysical observations of distant sources at X-ray and gamma-ray energies can be used to constrain the parameter space of ALPs. The propagation of high energy photons above a critical energy in the galactic and intergalactic fields may distort the intrinsic source spectra due to partial conversion of photons into ALPs or axions [23,24]. The comparison of observed and intrinsic source spectra gives indication for production of axions and can be used to understand the parameter space of ALPs. The observations of blazars during activities at X-ray and gamma-ray energies are important astrophysical events for searching the signature of ALPs in high energy astrophysics through photon-ALP oscillation. However, a deep understanding of the intergalactic magnetic field is equally important for observing the effects of photon-ALP oscillation on the gamma-ray spectra of blazars. The photon-ALP oscillation has potential effect on the opacity of the Universe to high energy gamma-rays coming from sources at cosmological distances. Observation of hard gamma-ray spectra from distant sources with ground based Cherenkov telescopes gives evidence for new physics involving photon-ALP oscillations. Observations of neutrino burst from supernovae are useful in constraining the properties of axions significantly [25].

# 4 Implications of ALPs

The existence of particles belonging to axion class in the real world can only be proven from their actual detection in the laboratory or in the astrophysical observations. The high energy astrophysical experiments along with the cosmological considerations are also useful in constraining the parameter space of the axions. The implications of detection of the axion like particles in different fields of the physics are discussed below.

## 4.1 Astrophysics

Very High Energy (VHE, E > 50 GeV) travelling cosmological distances through galactic and intergalactic magnetic fields show observational effects for the existence of ALPs. The opacity of Universe to VHE gamma-ray photons emitted from the extragalactic sources at high redshifts is drastically reduced due to conversion of photons into ALPs. Observations of VHE gamma-rays from the astrophysical sources like blazars and radio galaxies by the present and future generation ground based Cherenkov telescopes are very important to probe and constrain the properties of ALPs [26]. X-ray observations from active galactic nuclei are observed to most sensitive probes for photon-ALP conversion and detection of low mass axions. Improvements in the existing theoretical tools for analysing the photon-ALP conversion using astrophysical observations will lead to better the sensitivity for detection of axions. The axion-hadron coupling can be studied using low energy interactions at the core of supernovae [3]. Cooling rates of white dwarfs can be probed by studying the axion-electron interactions in their electron degenerate cores. Simulation studies of white dwarfs with and without axions in their cores are used to estimate the energy loss rate in such systems [27]. Therefore, the astrophysical observations using ground and space based experiments are important to search for ALPs and to probe the new physics beyond the SM of particle physics.

## 4.2 Cosmology

ALPs are expected to play a very important role in cosmology to explain inflation, cosmological constant problem, origin of cosmological initial conditions and dark matter. For QCD axions to be observable in cosmology, their contribution to the total energy density of the Universe should be significant. During the early times of Universe evolution, the axions are expected to have contribution to the vacuum energy. Therefore, the particles of axions class are important to explain the cosmological quantities like dark energy and inflation which is the period of accelerated expansion of the early Universe. Axions are also very crucial in solving the cosmological constant problem which is one of the important unsolved problems in modern cosmology. According to axion quintessence model, the potential energy of light mass axions drives accelerated expansion of the Universe and provides an effective cosmological constant [7]. Axions in the form of so called axion stars are promising candidates for dark matter [28]. Small cluster of axions are thought to be produced during the QCD phase transition in the early Universe which have formed compact coherent states of axions by condensation and gravitational cooling. The gravitationally bound states of axions are referred to as axion stars in modern cosmology. The mass of cosmological axions have been constrained between $10^{-6}$ eV and $10^{-3}$ eV which corresponds to the photons in MHz-GHz frequency band.

**4.3 Particle Physics**

Physical processes involving strong interactions indicate that the discrete symmetries based on charge conjugate (C) and parity (P) are good natural symmetries. Therefore, these symmetries must be respected in all theories of strong interactions. However, CP symmetry in strong interactions is observed to be violated in QCD due to non-zero vacuum angle (phase of the strong interaction). This is referred to as the famous strong CP problem in particle physics. Peccei and Quinn proposed that the vacuum angle can be associated with a phase by using a symmetry in the strong interactions [1]. This phase in PQ symmetry is described as the pseudo-scalar Goldstone boson which is known as axions. Involving the existence of this hypothetical particle, the PQ symmetry in QCD provides solution to the strong CP problem. The origin and connection of PQ symmetry with other aspects of particle physics are not clearly understood. It has been proposed that the PQ symmetry originates from the flavour symmetries which are used to understand the fermion masses and mixing [29]. The axions are related to the breaking of flavour symmetries. The discrete flavour symmetries are combined with the unified gauge theories (where leptons and quarks in SM are similar) to solve the strong CP problem involving quarks [30]. The neutrino masses and mixing parameters can also be explained by using discrete flavour symmetries. Therefore, detection of axions will help in the realization of PQ symmetry in nature and other related physical phenomenon in particle physics. The detection of axions will also lead to the identification of weakly interaction slim particles (WISPs) at new energy scales in particle physics.

# 5 Conclusions

Axions and ALPs are very useful in resolving a number of unsolved problems in astro-particle physics. The QCD axions solve the strong CP problem by settling down at the minima of the potential acquired from the non-perturbative effects. They can be embedded in the several extensions of SM of particle physics at high energy scales through different symmetries and fields. The axions in the form of axion stars are assumed to be mainly distributed as dark matter in the Universe. The QCD axions are considered as attractive candidates for dark matter due to their coupling to the standard model particles and well constrained parameter space. The natural sources for axions are the Sun and Big Bang from the early Universe.

Helioscopes (solar axion detector) and Haloscopes (dark matter halo axions) are supposed to be more sensitive axion detectors than the laboratory experiments like interferometry and laser shining through a wall because of high fluxes of natural axions. Several astrophysical observations in X-ray and gamma-ray energy bands can constrain the parameter space of axios better than other experiments. Future generation experiments in high energy astrophysics like Cherenkov Telescope Array (CTA) and in particle physics have great potential to probe the existence of axions and ALPs and to constrain their parameter space.